\documentclass[epj,twocolumn]{webofc}
\usepackage[varg]{txfonts}   
\usepackage{graphicx}
\woctitle{TRANSVERSITY 2014}
\begin{document}
\title{Transverse Single Spin Asymmetries in Hadronic Interactions}
%
%
\subtitle{An Experimental Overview and Outlook}

\author{L.C. Bland}

\institute{Brookhaven National Laboratory, Upton, New York  (USA)}

\abstract{
Transverse single-spin asymmetries (SSA) are expected to be small in
perturbative QCD because of the chiral nature of the theory.
Experiment shows there are large transverse SSA for particles produced
in special kinematics.  This contribution reviews the experimental
situation and provides an outlook for future measurements.}

\maketitle
\section{Introduction}
\label{intro}
We now agree that Quantum Chromodynamics (QCD) is the theory of the
strong interaction.  QCD describes mesons and baryons as being
composed of color-charged quarks ($q$) and anti-quarks that interact
via the exchange of gluons ($g$). Two non-trivial aspects of QCD are
that the gluons carry color charge and that color is absolutely
confined into color-neutral objects.  These aspects make it
complicated to understand the structure of mesons and baryons, and
lead to emergent phenomena that are not readily evident from the QCD
Lagrangian.  The quest to understand how the proton gets its spin from
its constituents is one avenue to tackling the big question regarding
color confinement.

Since the up and down quarks are so light and QCD is a
vector gauge theory, we expect that helicity is essentially unchanged
at the $q\rightarrow qg$ vertex \cite{KPR78}, with the probability for
helicity flip being proportional to the quark mass.  Transverse
single-spin asymmetries (SSA) are an azimuthal modulation of 
particles that can be observed either from decay or via spin-dependent particle
production.  Transverse SSA requires helicity flip, so are expected to
be small. Experiment observes large transverse SSA for particles
produced via the strong interaction in particular kinematics at
collision energies where the hadroproduction is described by
next-to-leading order (NLO) perturbative QCD (pQCD) calculations.

Spin-orbit correlations and $qg$ correlations are two suggestions by
theory why transverse SSA are so large.  Transverse momentum ($k_T$)
can be correlated with the spin of either the quark or hadron.  This
$k_T$ can be either in the initial state \cite{Si90} (Sivers effect)
or in the fragmentation of partons into hadrons \cite{Co93} (Collins
effect). An issue for the Sivers effect is that factorization theorems
have not been proven for the use of $k_T$-dependent distribution
functions to describe inclusive particle production in hadronic
interactions, except in the case of Drell-Yan production.
Factorization is used for collinear calculations \cite{QS91} that use $qg$
correlators \cite{ET82}.  The $qg$ correlators can appear in the
initial state or in the fragmentation, but are collinear so do not
involve $k_T$.  Explicit relations between initial-state $qg$
correlators and $k_T$ moments of the Sivers function have been found \cite{JQVY06}.
The Sivers function is important to understand because it can provide
new insight into the structure of the proton, regarding the role of
orbital motion of the confined partons \cite{Bu06,BR11}, although model independent
connections have not been found.

\begin{figure}[h]
   \centering
   \includegraphics[width=0.50\textwidth,clip]{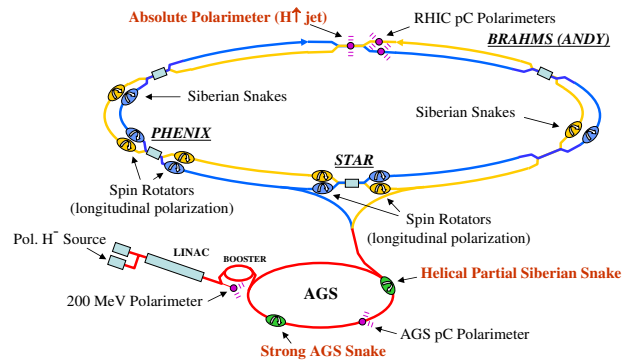}
   \caption{Schematic of RHIC as a polarized proton collider.
   Polarization is produced at the source, and is preserved through
   the acceleration sequence using Siberian Snake magnets.  Each ring
   has two full snakes that each precess the polarization
   by $180^{\circ}$.  Beams are transversely polarized in the rings.
   Spin rotator magnets can precess the polarization to become
   longitudinal at STAR and PHENIX.  The 2 o'clock interaction region
   was originally for the BRAHMS experiment, and later for the A$_N$DY
   experiment.  Results from both are discussed below.} 
\label{RHIC_spin}         
\end{figure}

\begin{figure*}[t]
   \centering
   \includegraphics[width=0.36\textwidth,clip]{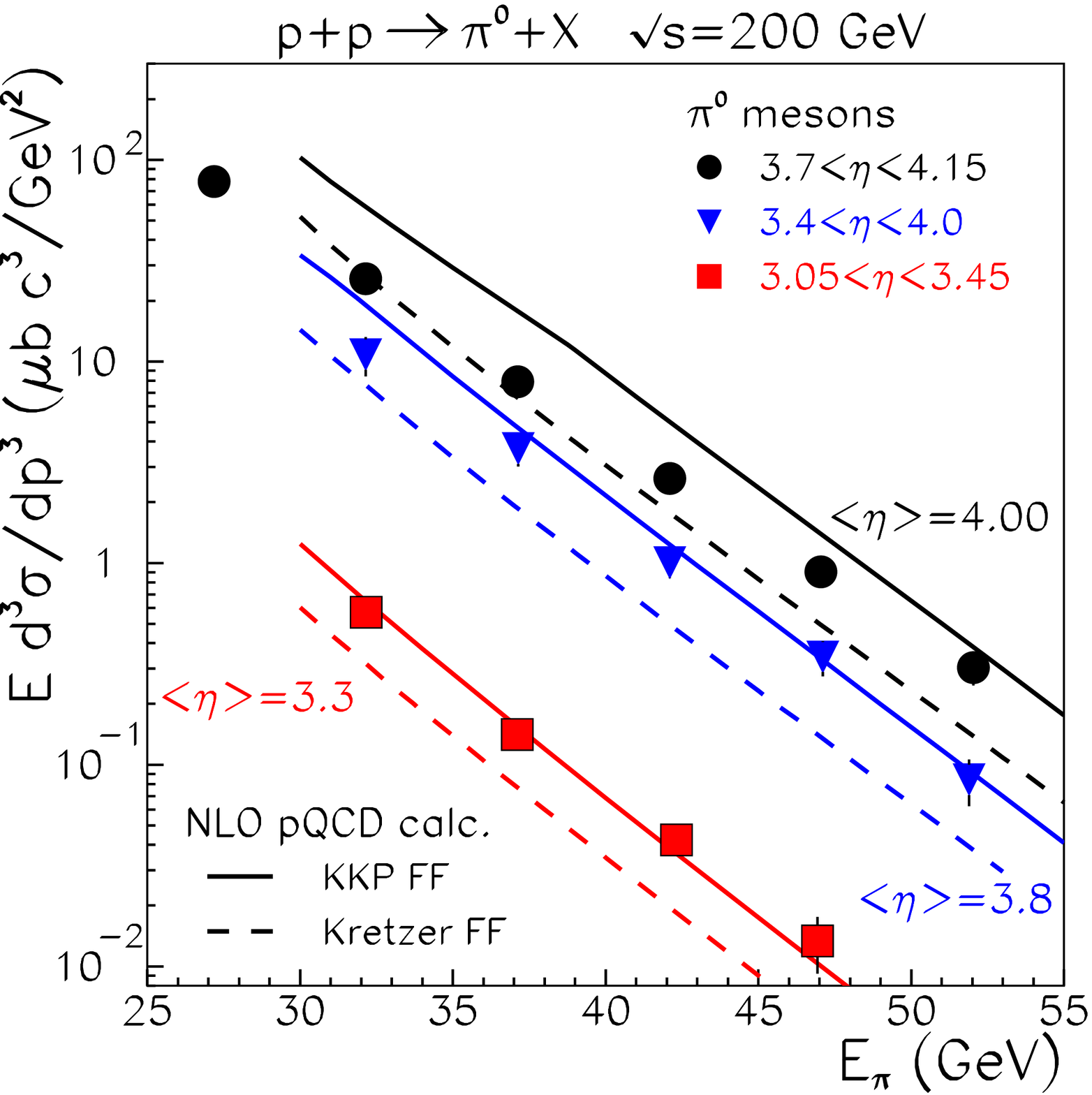}
   \includegraphics[width=0.44\textwidth,clip]{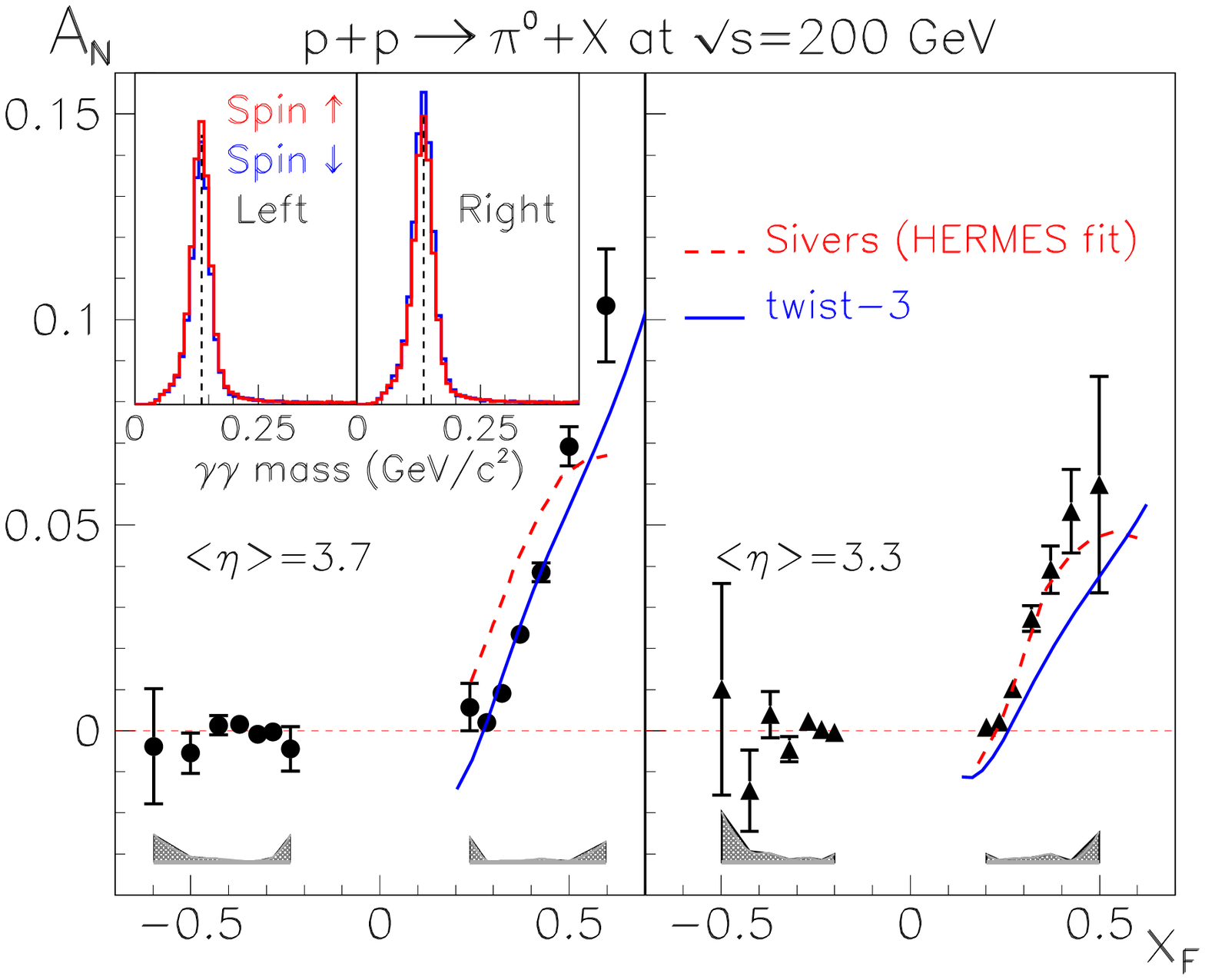}
   \caption{(Left) Cross sections for inclusive neutral pions produced at
   large $x_F$  \cite{STAR_FPD1} compared to NLO pQCD calculations.  (Right) Analyzing
   power for the inclusive production of neutral pions at large $x_F$ \cite{STAR_FPD2}, in
   comparison to calculations described in the text.} 
\label{STAR_FPD_figure}         
\end{figure*}

This contribution reviews recent experimental measurements of
transverse SSA in hadroproduction.  Operation of the Relativistic
Heavy Ion Collider (RHIC) at Brookhaven National Laboratory includes
polarized proton collisions, at center-of-mass energies spanning from
$62<\sqrt{s}<510$ GeV.  As the first and only polarized-proton
collider in the world, RHIC has provided significant new measurements
of transverse SSA.  Context of these new measurements is provided by
reference to older measurements at fixed-target facilities that
necessarily are at lower $\sqrt{s}$.  In addition, an outlook for
future measurements is provided.  Theoretical understanding of these
new measurements is still developing.  Given that understanding
emerges when experiment confronts theory, some discussion
will be provided.

\section{Transverse SSA measurements at RHIC}
\label{RHIC}
\subsection{RHIC spin}
\label{RHIC-spin}

Particle production at high energies typically involves accelerating
ion beams most commonly done with synchrotrons.  Preserving beam
polarization in high-energy synchrotrons is difficult because of many
intrinsic and imperfection resonances that can depolarize the beams.
Collisions of high-energy polarized beams at RHIC are made possible by
Siberian Snakes \cite{De78}.  RHIC realizes this concept by
superconducting helical dipole magnets that precess the polarization
vector by $180^{\circ}$ when the beam traverses the magnet, thereby
resulting in perturbations of the polarization vector about the stable
transverse direction as the polarized beams orbit the ring.  Each RHIC
ring has two Siberian Snake magnets.  Similar magnets at two of the
six interaction points (IP) can serve to precess transverse
polarization to become longitudinal for collisions, and then restore
transverse polarization after the IP.  Alternatively, transversely
polarized proton collisions can be studied.

It was recognized before the first polarized proton collision run that
local polarimeters would be required to measure whether spin-rotator
magnets were properly tuned to minimize polarization components that
were transverse to the beam momenta for the colliding beams.  Such
local polarimeters require identifying some sort of hadroproduction
from colliding beams that has non-zero transverse SSA.  Neutrons produced near $0^{\circ}$
were found to have a non-zero transverse SSA \cite{Fu07}.  The particle
multiplicity observed in beam-beam counters (scintillator annuli that
bracket the IP with acceptance near beam rapidity) was found to have
azimuthal modulations correlated with the transverse spin.  Finally,
neutral pion production at large rapidity was found to have a sizeable
transverse SSA \cite{STAR_FPD0}, although the production rate is such that its use as a
local polarimeter is limited.  Transverse SSA are important as a tool
for polarimetry.  Transverse SSA have intrinsic interest, as the rest
of this contribution will address.

The large RHIC experiments are at IP6 (STAR) and IP8 (PHENIX) in
Fig.~\ref{RHIC_spin}.  When RHIC began, IP2 had a traditional magnetic
spectrometer experiment with good particle identification (BRAHMS),
with one arm viewing large rapidity particle production.  More
recently, a forward calorimeter experiment (A$_N$DY, as proposed
in \cite{ANDY_PAC} and described in \cite{No11,Pe11})  was staged at IP2
for a brief time.  Both BRAHMS and A$_N$DY made transverse SSA
measurements, as discussed below.  The PHENIX and STAR experiments are most heavily
instrumented near midrapidity, although both experiments have
implemented forward electromagnetic calorimeters that enable access to
large-$x_F$ ($x_F=2p_L/\sqrt{s}$, is the Feynman scaling variable)
identified particle production.  Forward pion detectors at STAR were
made from lead glass, and viewed particles produced at $\sim
2.5<\eta<4.0$ through 1-m holes in the poletips of the 0.5 T solenoid
used to momentum analyze charged particles that are tracked through its time projection
chamber.  PHENIX implemented lead tungstate calorimeters (muon piston
calorimeter) that span $3.1<|\eta|<3.8$.

\begin{figure*}
   \centering
   \includegraphics[width=0.89\textwidth,clip]{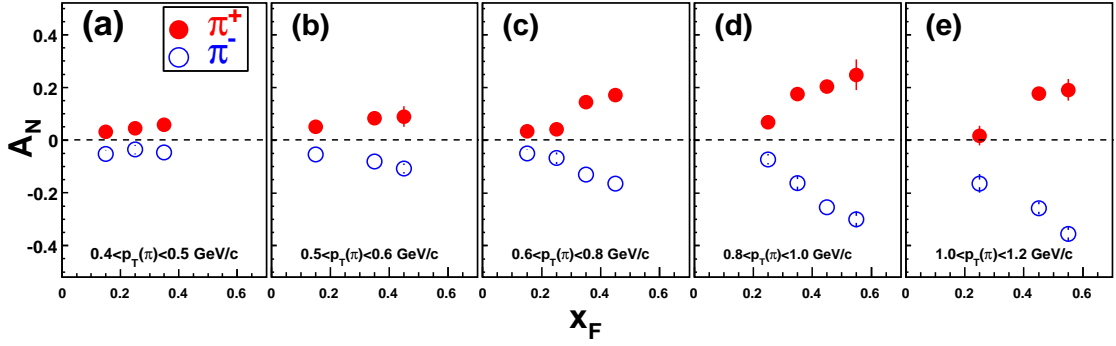}
   \caption{Analyzing power for $p^{\uparrow}+p\rightarrow
   \pi^{\pm}+X$ at $\sqrt{s}=62$ GeV \cite{BRAHMS62}.  Large $A_N$ is
   observed when the $\pi^{\pm}$ are produced in the forward direction.}
\label{BRAHMS}         
\end{figure*}

Another important concept for the early RHIC spin program was the
importance of measuring particle production cross sections for
comparison to NLO-pQCD calculations, done concurrently with measuring
spin asymmetries.  A primary motivation was to ensure that the spin
asymmetries were for properly reconstructed particles or ensembles.
Cross section comparisons to NLO-pQCD calculations are useful to
establish the applicability of theory to interpret the spin
observables.

Published work to date for transverse SSA are mostly for inclusive pion
production and for jets.

\subsection{Transverse SSA for inclusive pion production}
\label{pions}
Pions are prolifically produced in high energy hadroproduction.
Inclusive pion production is found \cite{STAR_FPD1} to have its cross
section described well by NLO pQCD at RHIC energies ($\sqrt{s}>62$
GeV), even for pions produced in the forward direction, as defined when
$x_F$ is sizeable.  There are non-zero transverse SSA for pion
production \cite{STAR_FPD2} at large rapidity
(Fig.~\ref{STAR_FPD_figure}), in the same kinematics where the
spin-averaged cross section is in agreement with NLO pQCD.  The
transverse SSA for particles produced from a transversely polarized
proton beam is
\begin{equation}
A_N=\frac{\sigma_{\uparrow}-\sigma_{\downarrow}}{\sigma_{\uparrow}+\sigma_{\downarrow}}.
\label{AN}
\end{equation}
This transverse SSA is called analyzing power, where $\sigma_{\uparrow
/ \downarrow}$ refers to the particle production cross section for
different directions of the beam polarization vector.

\begin{figure}
   \centering
   \includegraphics[width=0.42\textwidth,clip]{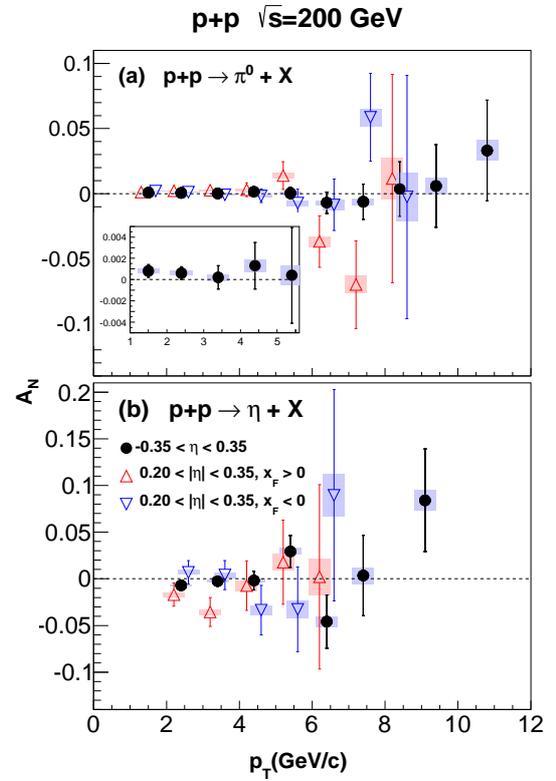}
   \caption{Analyzing power for $p^{\uparrow}+p\rightarrow
   \pi^0+X$ at $\sqrt{s}=200$ GeV \cite{PHENIX1}.  The $\pi^0$ are
   detected at midrapidity.  Also shown is $A_N$ for
   $\eta\rightarrow\gamma\gamma$ reconstructed at midrapidity.}
\label{PHENIX}         
\end{figure}

Operationally, $A_N$ requires measurement of integrated
luminosity for spin-up ($\uparrow$) and spin-down ($\downarrow$)
beams.  As well, the beams do not have all particles with spins
pointing in a particular direction, but instead are an ensemble of
particles having a polarization, $P_{beam}$, measured in independent counting
experiments.  For the data in Fig.~\ref{STAR_FPD_figure}, a carbon fiber was
inserted into the beams at regular times for each fill and the spin
dependence of recoil carbon ions was measured.  The momentum transfer
for the elastic scattering of polarized protons from carbon is in the
region where the Coulomb amplitude for the process interferes with the
nuclear amplitude.  This Coulomb-Nuclear Interference (CNI)
polarimeter is considered a relative polarimeter, because the
spin-dependence of the nuclear amplitude is not known a priori, unlike
for the Coulomb amplitude, where the spin dependence is determined
from the anomalous magnetic moment of the proton.  The effective
normalization of the CNI polarimeter is completed by having the
high-energy polarized proton beams scatter from a gas jet of hydrogen
atoms, where the protons in this jet are polarized \cite{Ok06}.  Identical
particle symmetries allow transfer of knowledge of the polarization of
the gas jet to polarization of the proton beam.  The counting rates
for elastic scattering from the polarized gas jet initially required
multiple fills of RHIC to get sufficient statistical precision on the
beam polarization.

\begin{figure*}
   \centering
   \includegraphics[width=0.37\textwidth,clip]{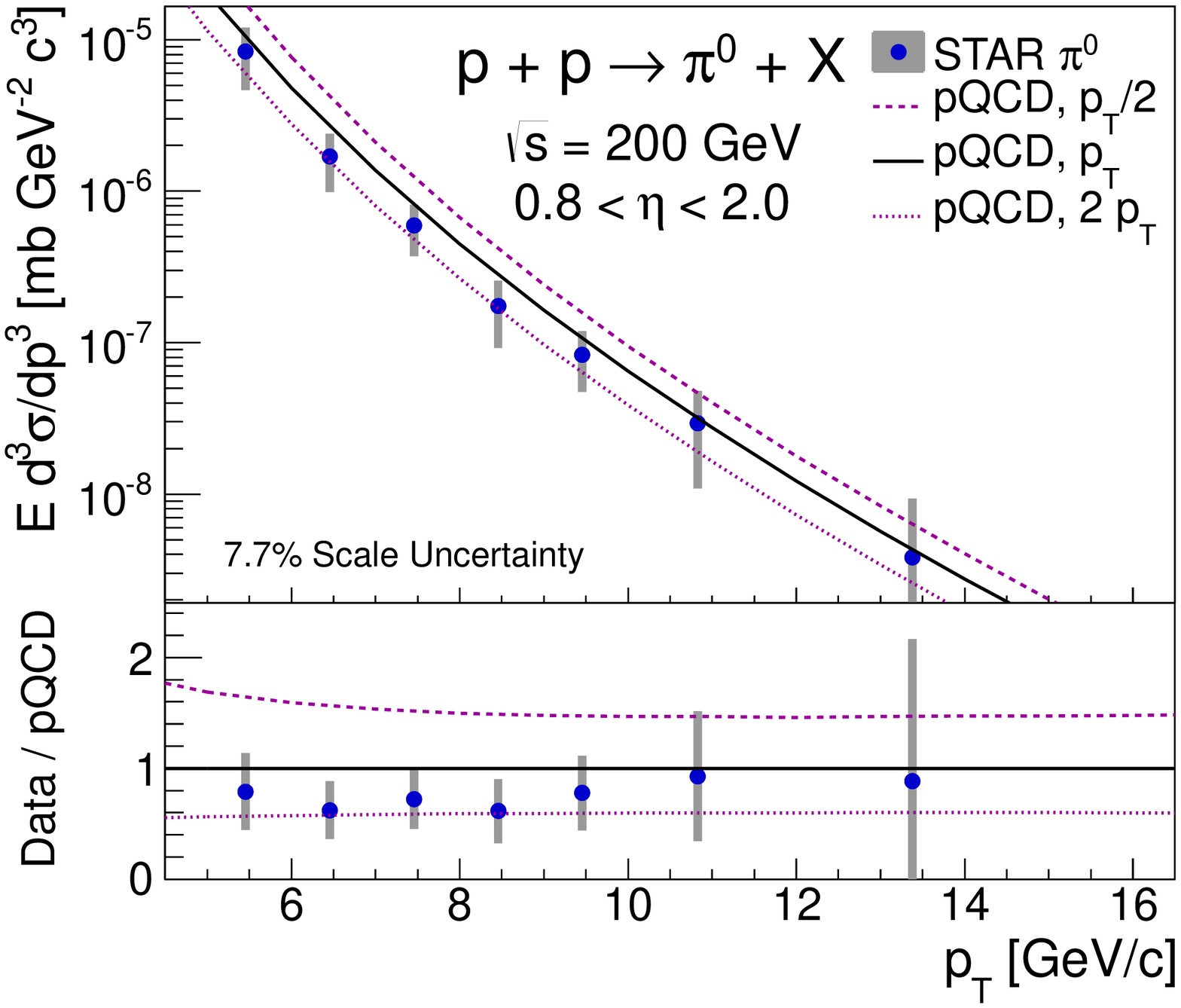}
   \includegraphics[width=0.48\textwidth,clip]{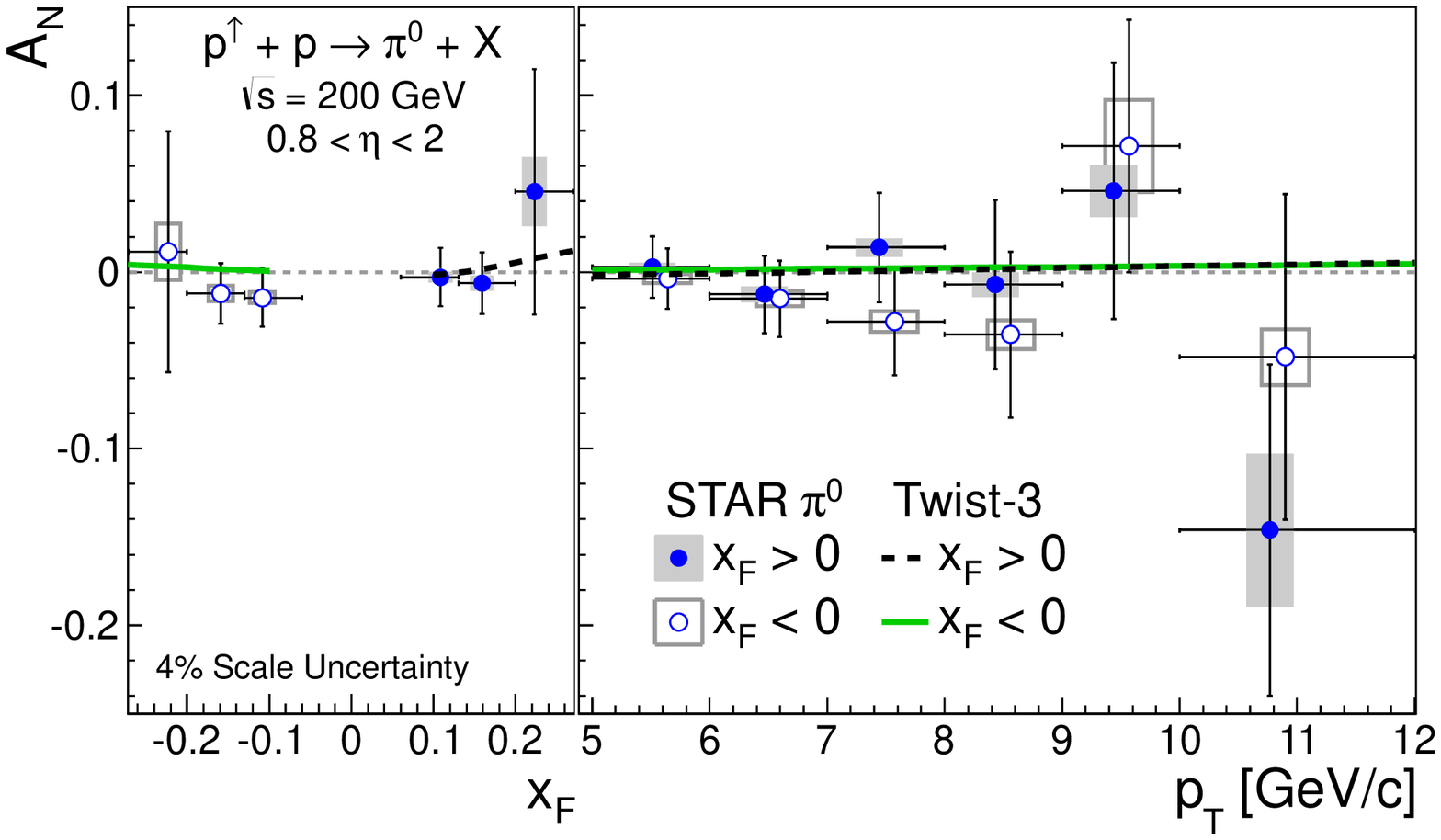}
   \caption{(Left) Cross sections (left) and $A_N$ (right) for inclusive neutral pions produced at
   mid-central rapidity ($0.8<\eta<2.0$) \cite{STAR_midcentral}.}
\label{STAR_eemc}         
\end{figure*}

$A_N$ can be measured as a left/right asymmetry of the particle
production from $p^{\uparrow}p$ collisions in the reaction plane, defined by by the momenta of the
beams and the produced particle.  This left/right asymmetry can be
non-zero when there is a component of the beam polarization
perpendicular to the reaction plane.  The convention is that $A_N>0$
when more particles are produced to the left when $P_{beam}$ is up. Such a measurement
requires only a single direction for the beam polarization and
knowledge of the acceptance of the left and right detectors.  Mirror
symmetrical calorimeter modules were used for the measurements in
Fig.~\ref{STAR_FPD_figure}. Since spin-up and spin-down polarizations were
both available, operationally
\begin{equation}
A_N=\frac{1}{P_{beam}}
    \frac{\sqrt{N_{\uparrow}^L N_{\downarrow}^R}-\sqrt{N_{\downarrow}^L N_{\uparrow}^R}}
         {\sqrt{N_{\uparrow}^L N_{\downarrow}^R}+\sqrt{N_{\downarrow}^L N_{\uparrow}^R}},
\label{cross ratio}
\end{equation}
where $N_{\uparrow / \downarrow}^L(R)$ refers to particle production
to the left (right) of the beam whose polarization magnitude is
$P_{beam}$ pointing up ($\uparrow$) or down ($\downarrow$).

$A_N$ was measured for $p^{\uparrow}+p\rightarrow\pi^{\pm}+X$ at
$\sqrt{s}=62$ GeV \cite{BRAHMS62} by the BRAHMS collaboration
(Fig.~\ref{BRAHMS}).  Charged pion production cross sections in these
same kinematics were found to agree with NLO pQCD, as for neutral pion
production, at $\sqrt{s}=200$ GeV \cite{BRAHMS2}.  Preliminary
results show similar agreement between charged pion cross sections and
NLO pQCD at $\sqrt{s}=62$ GeV \cite{BRAHMS3}.  BRAHMS was a
traditional magnetic spectrometer with particle identification, so
relied on concurrent measurement of spin-dependent integrated
luminosities to measure $A_N$ according to Eqn.~\ref{AN}.  $A_N$ at
large negative $x_F$ is also reported at $\sqrt{s}=62$ GeV, and found
to be consistent with zero.  Large positive $A_N$ is found for
$p^{\uparrow}+p\rightarrow K^{\pm}+X$.  The large positive $A_N$ for
$K^-$ production suggests a role played by the sea of $q\overline{q}$
pairs within the proton, if the transverse SSA is an initial-state effect.

\begin{figure}
   \centering
   \includegraphics[width=0.38\textwidth,clip]{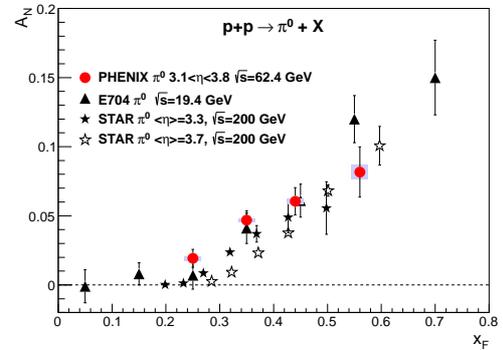}
   \caption{Comparison of the $\sqrt{s}$ dependence of large-$x_F$
   neutral pion production \cite{PHENIX1}.}
\label{PHENIX_roots}         
\end{figure}

Inclusive $\pi^0$ production has also been measured at mid-rapidity by
the PHENIX collaboration (Fig.~\ref{PHENIX}). $A_N$ is found to be
consistent with zero at mid rapidity \cite{PHENIX1}.  Particle
production cross sections in these rapidity intervals are found to be
consistent with NLO pQCD \cite{PHENIX_sigma}.  They also report $A_N$ for
$\eta\rightarrow\gamma\gamma$ at midrapidity, and find it too is
consistent with zero.  As will be discussed below, the midrapidity
measurements span the same $p_T$ range where $A_N(\pi^0)$ is large at
large $x_F$.

PHENIX has implemented a forward electromagnetic calorimeter (the muon
piston calorimeter).  They have reported \cite{PHENIX1} $A_N(\pi^0)$ at large $x_F$
for $p^{\uparrow}+p$ collisions at $\sqrt{s}=62$ GeV.  Their results
are found to be consistent (Fig.~\ref{PHENIX_roots}) with those from
Fig.~\ref{STAR_FPD_figure}, although at a lower $\sqrt{s}$.  Also
shown in Fig.~\ref{PHENIX_roots} are $A_N(\pi^0)$ measurements made by
the E704 collaboration at FermiLab.  E704 used a 200 GeV polarized
proton beam incident on a fixed target \cite{E704-1}.  The $\sqrt{s}$
dependence for $A_N$ will be discussed further below.

\begin{figure*}
   \centering
   \includegraphics[width=0.63\textwidth,clip]{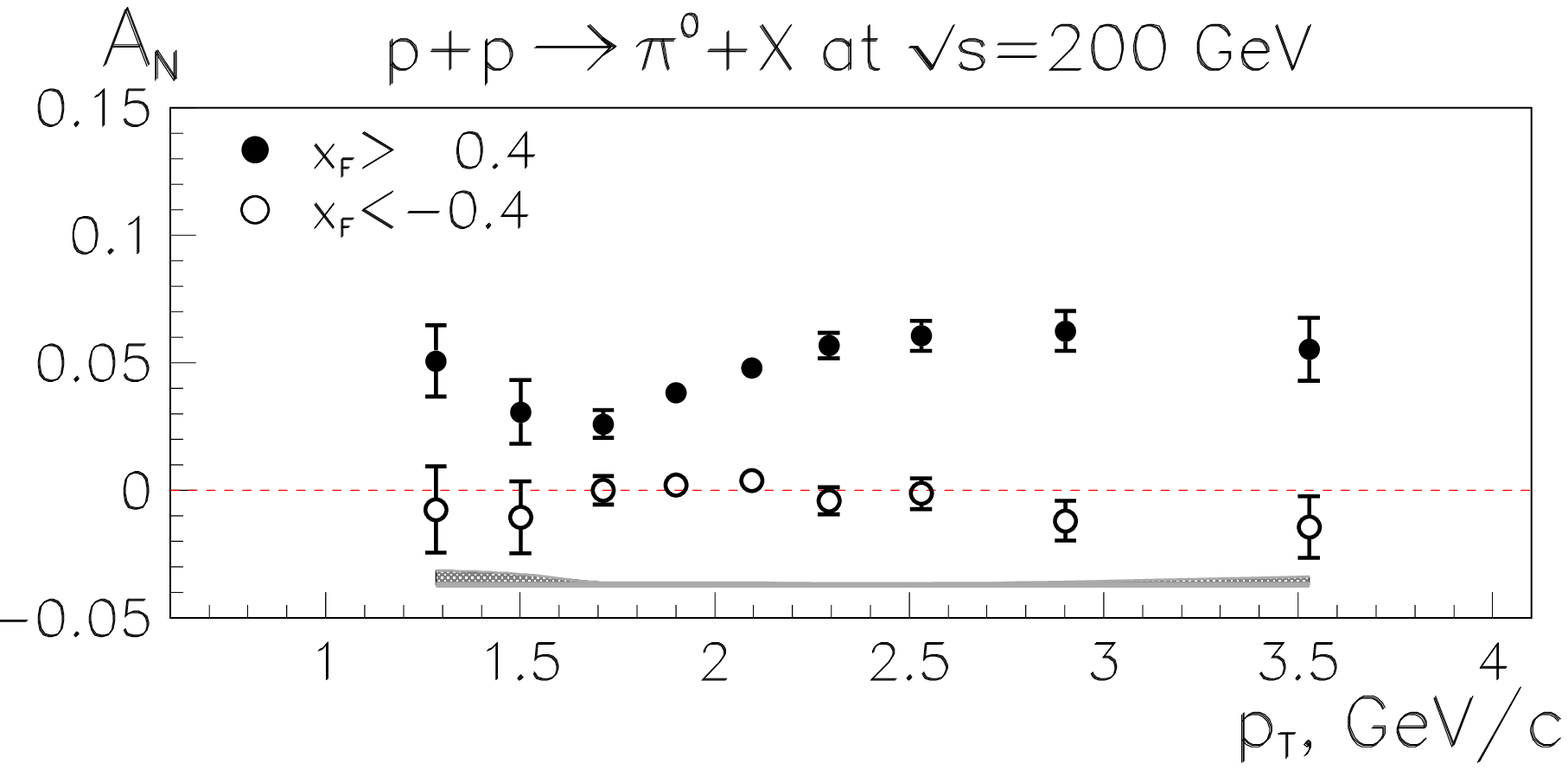}
   \includegraphics[width=0.35\textwidth,clip]{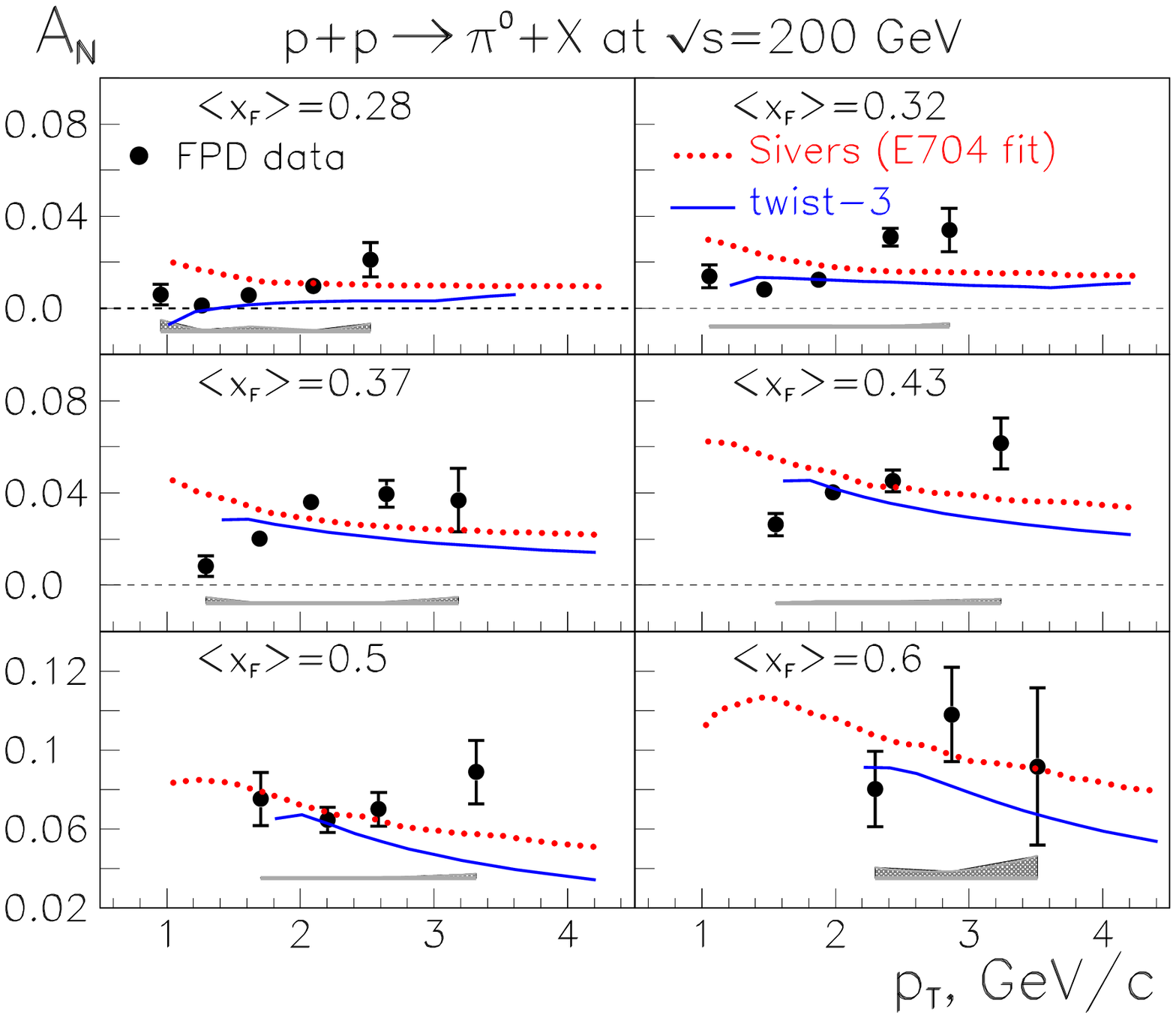}
   \caption{The $p_T$ dependence of $A_N$ for forward neutral pion
   production is shown.  (Left) $p_T$ dependence requiring a threshold
   value for $x_F$ and (right) fully separated $x_F,p_T$ dependencies
   both show a plateau of $A_N$ with increasing $p_T$ \cite{STAR_FPD1}.}
\label{STAR_FPD_pT}         
\end{figure*}

Neutral pion results have been reported by the STAR collaboration at
rapidities intermediate between the central and forward regions
($1<\eta<2$) \cite{STAR_midcentral}.  As for the central and forward rapidity
regions, cross sections at mid-central rapidity are consistent with NLO pQCD.  $A_N$ for
neutral pion production is consistent with zero (Fig.~\ref{STAR_eemc})
in this mid-central rapidity region.

Simple patterns are evident in the data.  Spin-averaged cross sections
are in agreement with NLO pQCD over a broad range of rapidity for
$\sqrt{s}>62$ GeV.  Transverse SSA are consistent with zero, except in the forward
direction.  A possible explanation for this is that the dynamics that gives
rise to $A_N$ involves valence quarks, that are not readily accessible
at midrapidity until one reaches $p_T>10$ GeV/c.

\subsection{$p_T$ dependence of transverse SSA for forward neutral pion production}

In the forward region, it is possible to disentangle $p_T$ and $x_F$
dependences, because both longitudinal and transverse momentum
components can be large.  Extensions to pQCD that are model-dependent
applications of TMD or application of $qg$
correlators in a collinear framework both naively expect that
$A_N\propto 1/p_T$, for sufficiently large $p_T$.

For the $p_T$ range measured to date, $A_N$ is found to rise with
increasing $p_T$ as it must since there is no distinction between left
and right at $p_T=0$.  The transverse SSA stays constant at high
$p_T$, in the range accessible by experiment
(Fig.~\ref{STAR_FPD_pT}).  Preliminary results have extended the $p_T$
range for measurements of $A_N$ in neutral pion production out to
$\sim10$ GeV/c for $p^{\uparrow}+p$ collisions at $\sqrt{s}=500$ GeV
\cite{heppelmann}.  

The basic form of the $p_T$ dependence of $A_N$ for $p^{\uparrow} p
\rightarrow\pi X$ is reminiscent of other transverse SSA phenomena, as
measured in fixed-target experiments.  Such $p_T$ dependencies are
observed for the induced polarization of hyperons in unpolarized
hadroproduction: {\it e.g.} $pp\rightarrow\Lambda^{\uparrow}X$.  When
the $\Lambda$ is produced at moderate to large $x_F$ it has its spin
preferentially (anti) aligned transverse to the production plane.  At
fixed $x_F$, the induced polarization magnitude increases with $p_T$
to a plateau, and then persists to the highest $p_T$ values accessible
by experiment \cite{Lu89}.  Although phenomenological treatments can
explain the $p_T$ dependence of $p^{\uparrow}p\rightarrow\pi X$
\cite{An13,KKMP14}, these quantifications do not provide physical insight 
into this behavior.

\begin{figure*}
   \centering
   \includegraphics[width=0.90\textwidth,clip]{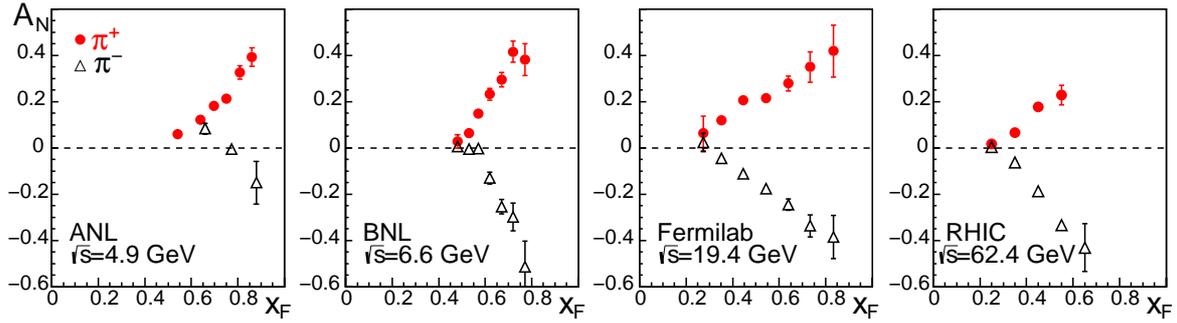}
   \caption{$A_N$ for $p^{\uparrow}p \rightarrow \pi^{\pm}X$ showing
   that the characteristic $x_F$ dependence spans a broad range of
   $\sqrt{s}$, from \cite{ABHM12} and measurements from \cite{Kl76,Al02,Ad91,BRAHMS62}.}
\label{an_roots}         
\end{figure*}

\subsection{Discussion}
Theory has worked to explain the large transverse SSA for
$p^{\uparrow}+p\rightarrow\pi+X$ in the forward direction.  Those
explanations are constrained by measurements of transverse SSA
in semi-inclusive deep inelastic scattering (SIDIS) for both the Sivers \cite{HERMES-2,COMPASS-1} and
Collins effects \cite{HERMES-1,COMPASS-2}, and in $e^+e^-$ collisions
for the Collins effect  \cite{Belle}.
Questions about factorization in SIDIS and $e^+e^-$ collisions have
been settled.  Factorized forms for these transverse SSA exist despite
the presence of final-state interactions in SIDIS that are required by
gauge invariance, and are required for there to be transverse SSA.
Factorized forms also exist for Drell-Yan (DY) production via
$p^{\uparrow} p \rightarrow\gamma^*X$, or for generalized DY production
of vector gauge bosons.  For DY production, theory predicts the Sivers
function will change sign relative to SIDIS because the attractive
final-state interaction in the latter \cite{BHS02} become a repulsive initial-state
interaction in the former \cite{Co02}.  As will be discussed below, this
prediction awaits an experimental test.  Complications for
$p^{\uparrow} p \rightarrow\pi X$ are that a mix of initial-state and
final-state interactions are in general possible and that factorization
for TMD distribution functions has not been proved.

One theoretical approach has been to proceed with use of TMD
distribution and fragmentation functions, despite not having proven
factorization for the hadroproduction of hadrons.  This approach will
be called generalized parton model (GPM) phenomenology in the
following discussion.  Another approach has been to do phenomenology
using $qg$ correlators, in a collinear twist-3 factorized
framework. It was generally accepted in the community that the {\it
soft-gluon pole} correlator was domninant.  This correlator is related
to the $k_T$ moment of the Sivers function \cite{JQVY06}.

A fundamental difficulty is that TMD distribution and fragmentation
functions are objects with two scales.  In SIDIS, these two scales are
the virtuality of the photon ($Q^2$) and the transverse momentum of
the observed hadron ($p_T$).  For $p^{\uparrow} p \rightarrow\pi X$ there
is only a single scale, given by the $p_T$ of the observed $\pi$.
This single scale does not provide access to either the magnitude of
the TMD transverse momentum ($k_T$) or to whether it acts in the
intial-state (via the Sivers effect) or the final-state (via the
Collins effect).

Theoretical calculations in Fig.~\ref{STAR_FPD_figure} are GPM
calculations \cite{Bo08} that fit Sivers moments in SIDIS
\cite{HERMES-2,COMPASS-1} and twist-3 calculations that use
initial-state $qg$ correlators and soft-gluon pole dominance, fitted
to $p^{\uparrow}p\rightarrow \pi X$ data only \cite{KQVY06}.
Compatibility of calculations of $p^{\uparrow}p \rightarrow \pi X$ in
the twist-3 approach and extractions of the Sivers function from SIDIS
has been examined.  Because of the expected dominance of initial-state
interactions for $p^{\uparrow}p \rightarrow \pi X$, $A_N$
is found to be opposite in sign to that of the transverse SSA for SIDIS \cite{KQVY11},
using initial-state $qg$ correlators and the relation to moments of
the Sivers function \cite{JQVY06}.  This sign mismatch has prompted
speculations that the Sivers function may have a node.  Another
solution was presented at this workshop \cite{Pitonyak}.  Namely, the
initial expectation that the {\it soft-gluon pole} dominates for
$p^{\uparrow}p \rightarrow \pi X$ is no longer considered valid
\cite{PM13,KKMP14}.  A $qg$ correlator in fragmentation, that is not
related via a $k_T$ moment to the Collins function, is now believed to
be the dominant contribution to $A_N$.  Phenomenology in this new
ansatz can provide a global explanation of SIDIS and
$p^{\uparrow}p\rightarrow \pi X$ data.  Sivers contributions are still
found by twist-3 phenomenology, but they are smaller than initial
estimates.  The {\it soft-gluon pole} $qg$ correlators are now negative, thereby cancelling
large positive contributions to $A_N$ from $qg$ correlators in
fragmenetation.

GPM phenomenology still expects that the Sivers effect dominates $A_N$
from $p^{\uparrow}p \rightarrow \pi X$.  The issue for the GPM remains
factorization, as the proponents have pointed out.

No theory to date provides an explanation for the persistance of
transverse SSA in $p^{\uparrow}p \rightarrow \pi^{\pm}X$ over a very
broad range of $\sqrt{s}$ (Fig.~\ref{an_roots}).  The transverse SSA
at $\sqrt{s}<20$ GeV most likely requires an explanation in terms of
mesons and baryons.

It would also be interesting to see the prediction for the Collins
angle distribution of the tranverse SSA for a $\pi^0$ within a jet for
the final-state twist-3 $qg$ correlator now thought to be the dominant
contribution to $A_N$ for $p^{\uparrow}p\rightarrow\pi X$.  There are
preliminary data \cite{Po11}, that still require determination of the
jet-energy scale, that show no dependence of the transverse SSA on the
Collins angle.  Determination of the jet axis and measurement of the
spin-correlated azimuthal modulation of the $\pi$ yield about this
axis is expected to have small Collins contributions within the GPM
\cite{DMP13}.  Azimuthal modulations of the $\pi$ yield within the jet
is a two-scale problem analogous to SIDIS, in that the jet $p_T$ and
the pion $k_T$ within the jet are both measured.

\begin{figure*}[th!]
   \centering
   \includegraphics[width=0.42\textwidth,clip]{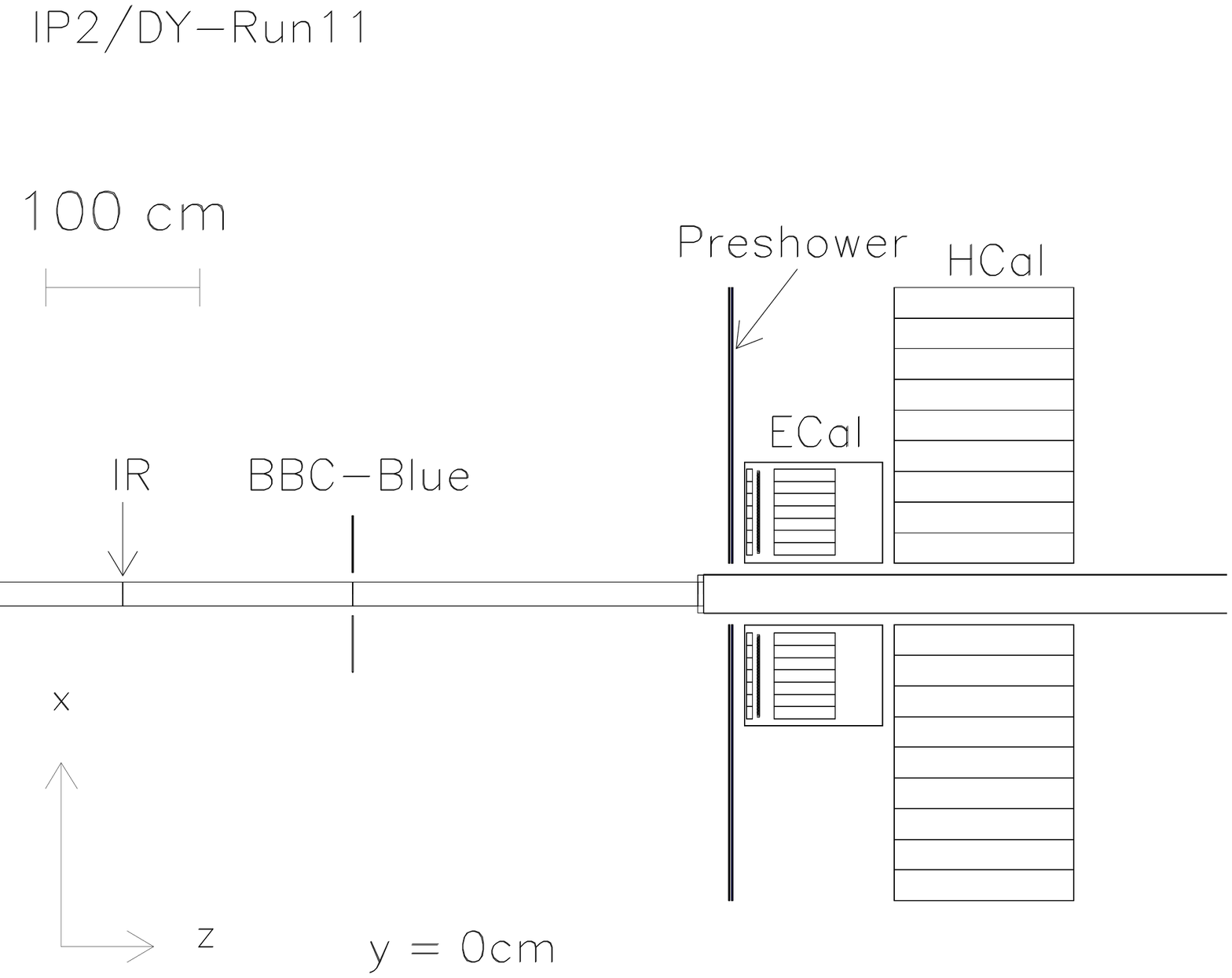}
   \includegraphics[width=0.37\textwidth,clip]{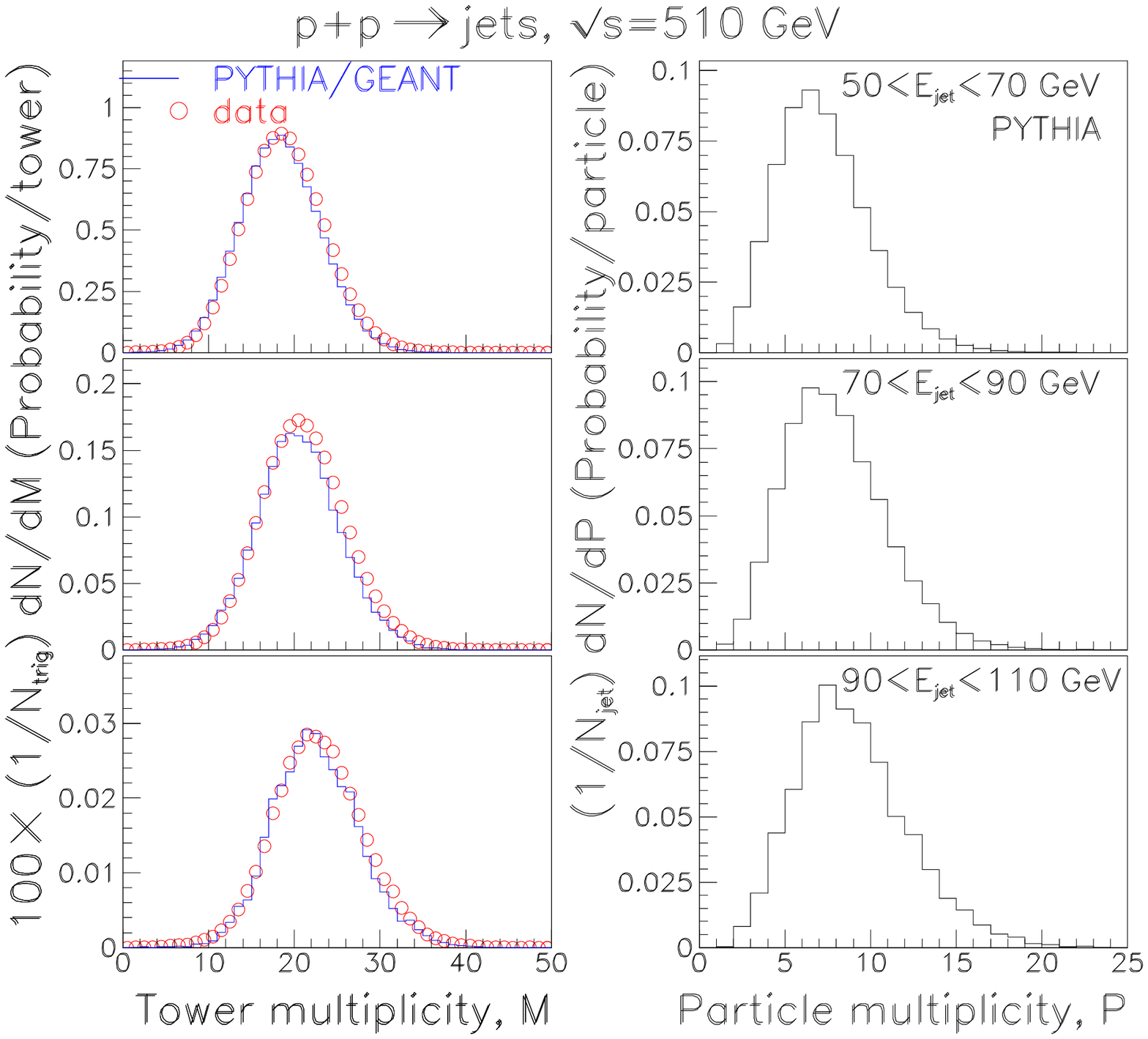}
   \includegraphics[width=0.20\textwidth,clip]{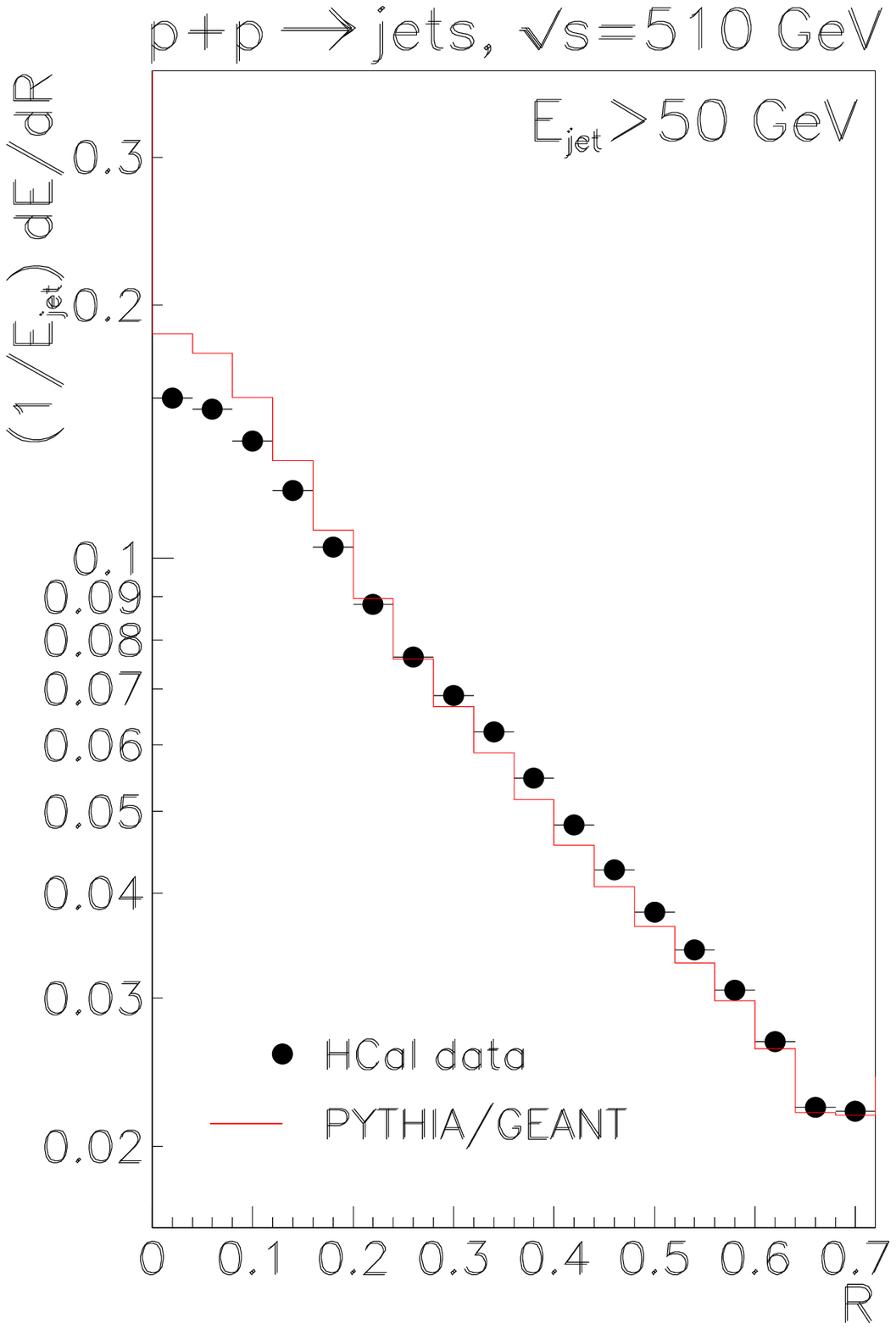}
   \caption{(left) Schematic of apparatus used for $p^{\uparrow}p
   \rightarrow$ jet$+X$ \cite{AnDY1}; (middle) multiplicities from the
   anti-k$_T$ jet-finding algorithm for (left panel) the HCal response
   in data and full simulation and (right panel) for particles as
   generated by PYTHIA \cite{PYTHIA6222}; (right) distribution of energy
   with respect to jet axis.}
\label{jet_figs}         
\end{figure*}

The question then is {\it where does this leave us}?  I think the
answer is that $p^{\uparrow}p \rightarrow \pi X$ has stimulated the
community to understand why such large transverse SSA exist, despite
the chiral properties of QCD.  Consequently, we are on the cusp of having a much
richer understanding of the structure of the proton, which remains the
quest.  To test that understanding, transverse SSA in $p^{\uparrow}p$
are important to establish a form of {\it universality} of the
phenomena.  The task at hand for $p^{\uparrow}p$ collisions is to go beyond inclusive $\pi$
production to jets, direct photons and DY production.  In the
remainder of this contribution, these first steps are discussed.  An
outlook to the future is then provided.

\subsection{Transverse SSA for inclusive jet production}
\label{jets}

Operations of RHIC for polarized proton collisions at $\sqrt{s}=500$
GeV were even more challenging than operations at $\sqrt{s}=200$ GeV,
because of the requirements on the accelerator to preserve polarization to
higher beam energies.  The primary focus of $\sqrt{s}=500$ GeV
collisions was to measure the parity-violating, longitudinal
single-spin asymmetry for the production of $W^{\pm}$ bosons.  A
proposal was put forth \cite{ANDY_PAC} to concurrently pursue first measurement of
$A_N$ for forward DY production to test the sign-change prediction.
The first stage of the apparatus required for that measurement was
staged at IP 2, in the hall originally used by the
BRAHMS collaboration.  That first stage apparatus used left/right
symmetric hadron calorimeter modules, as shown in Fig.~\ref{jet_figs}
The apparatus was ideal for measurements of $p^{\uparrow}p \rightarrow
$ jet$+X$, as discussed below.

There are many preconceptions about forward hadroproduction, and
extentions from inclusive $\pi$ production to jets immediately raises
the question about what we mean by jets.  To a theorist, a jet is a
scattered parton.  Factorized approaches ignore the
couplings of hard-scattered partons to spectator partons that are
required by gauge invariance, by the definition of factorization.  In
models, such as the string model, these couplings give rise to
initial-state and final-state parton showers which also serve to
complicate the definition of a jet.  Despite these complexities, we
proceed.

A jet is operationally defined as a pattern of energy deposition in a
localized region of $\eta-\phi$ space.  Multiple 
algorithms exist to recognize such patterns.  The favored algorithm is
the anti-$k_T$ method \cite{CSS08}, where all pairings of granular objects in
$\eta-\phi$ space are considered in the construction of a jet
pattern.  The granularity in this case is provided by the cells in the
hadron calorimeter.  We use $R=0.7$ for the jet finding, corresponding
to the jet-cone radius in $\eta-\phi$ space.  The mid-point cone jet
finder has also been used, with similar results \cite{No12}.

The result from applying the anti-$k_T$ algorithm to the calibrated
response of the modular calorimeters is an object that coincides with
our understanding of a jet (Fig.~\ref{jet_figs}), albeit with less
particle multiplicity than is observed at mid rapidity because the
transverse momentum of the jet is small and $p_T$ is generally taken
as the scaling variable for QCD treatments.  Forward jets have
multiplicities that match those from jet studies in $pp$ collisions in
fixed target experiments \cite{BCA}.  The distribution of energy as a function of
the distance from the jet axis in $\eta-\phi$ space coincides with our
expectations of what a jet should look like.

\begin{figure}
   \centering
   \includegraphics[width=0.34\textwidth,clip]{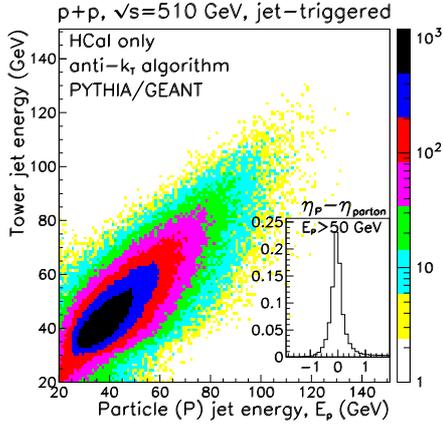}
   \caption{Comparison of results from the anti-k$_T$ algorithm
   applied to full PYTHIA + GEANT simulations versus events generated
   by PYTHIA \cite{PYTHIA6222}.  The inset shows the directional match between particle
   jets and hard-scattered partons, and results in an 82\% match when
   $|\Delta\eta|,|\Delta\phi|<0.8$.}
\label{jet_cal}         
\end{figure}

\begin{figure*}
   \centering
   \includegraphics[width=0.42\textwidth,clip]{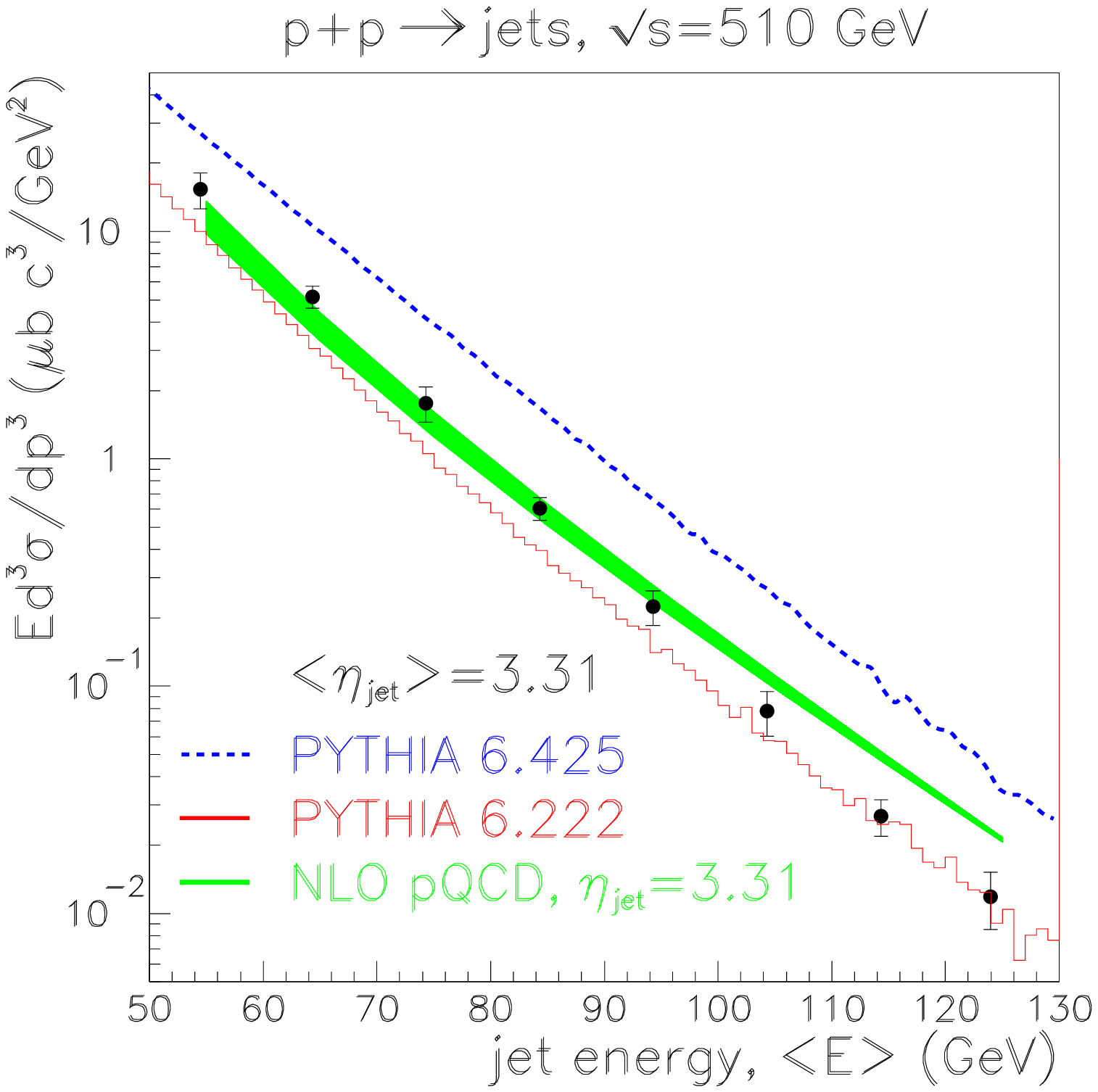}
   \includegraphics[width=0.44\textwidth,clip]{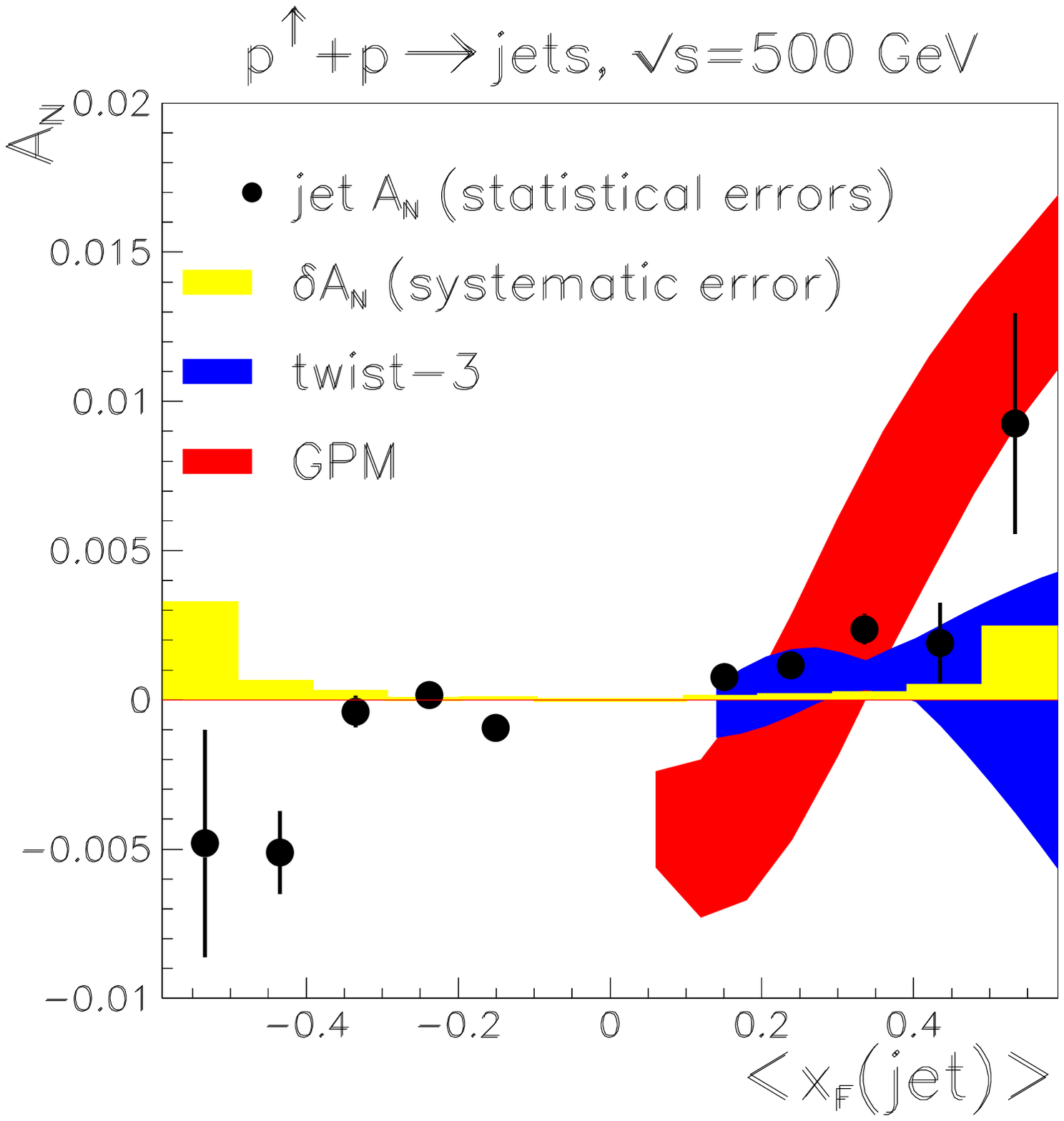}
   \caption{(left) Cross section for forward jet production in $pp$
   collisions at $\sqrt{s}=510$ GeV, in comparison to NLO pQCD and
   PYTHIA; (right) $A_N$ for forward jet production, in comparison to
   calculations that fit the Sivers effect for SIDIS.}
\label{jet_results}         
\end{figure*}

One note here:  calibrating the response of the calorimeters is the
essential and non-trivial step.  The calibrations are done by applying
particle finding algorithms, as described elsewhere \cite{Bl13}.  Both
electromagnetic and hadronic responses have been calibrated.  The
reconstructed jets are compared against particle jets reconstructed
from PYTHIA (Fig.~\ref{jet_cal}).

Jet finding integrates over hadronic fragments.  There are at least
two signficant implications: non-zero transverse SSA can only arise
from initial-state spin-correlated $k_T$ or initial-state $qg$
correlations and given the mirror symmetry ($A_N(\pi^+)\approx
-A_N(\pi^-)$), the naive expectation is that the analyzing power for
jets should be small.

Results for the forward jet cross section and $A_N$ are shown in
Fig.~\ref{jet_results}.  Cross sections are found to be in fair
agreement with NLO pQCD calculations \cite{Mu12}, as for forward $\pi$
production at $\sqrt{s}>62$ GeV.  Also shown are comparisons to
particle-jet results from two versions of PYTHIA.  PYTHIA 6.222 \cite{PYTHIA6222} is the
last version prior to tunings to explain underlying event
contributions for midrapidity particle production at the Tevatron and
PYTHIA 6.425 \cite{PYTHIA6425} includes first tunings, as done for preparation for the
LHC.  Forward particle production was not a criteria for tunings that
were made, and was impacted by those tunings.  This is particularly
relevant for QCD backgrounds to forward DY production, discussed
below.

The forward jet $A_N$ is non-zero for $x_F>0$.  Collins contributions
are not present, to the extent that the jet finding integrates over
all fragments, as suggested it does from comparions of particle jet
results to hard scattered partons (Fig.~\ref{jet_cal}).  Consequently,
in the TMD framework, $A_N$ for forward jet production arises only
from the Sivers effect.  The anticipated cancellation of $\pi^+$ and
$\pi^-$ contributions is observed, in that the magnitude of the jet
$A_N$ is small.  Comparisons to theory that fit the Sivers function
deduced from SIDIS are shown in Fig.~\ref{jet_results}.  The
generalized parton model (GPM) assumes factorization, and uses the
Sivers function from SIDIS directly in their calculation \cite{An13}.  Error bands
on the calculation reflect uncertainties in the Sivers functions from
SIDIS.  The twist-3 calculation uses soft-gluon pole $qg$ correlators
constrained to $k_T$ moments of the Sivers function \cite{Ga13}.  This calculation
has been cited as evidence of the color-charge reinteractions that
give rise to the predicted sign change from SIDIS to DY.

Mention should be made of $A_N$ for $x_F<0$.  The $p^{\uparrow}p
\rightarrow \pi X$ results generally have $A_N$ consistent with zero
at negative $x_F$.  The jet $A_N$ does have a negative analyzing power
with a $\sim 3.5$ sigma signficance at $x_F\approx -0.4$.  As we
heard at this workshop \cite{Koike}, tri-gluon correlators do predict negative
analyzing power for jet production at large negative $x_F$.  For the
forward jet production, the beam with $p_z$ opposite to that of the
detected jet is a source of low-$x$ partons, in a conventional
$2\rightarrow 2$ partonic scattering picture for the particle
production.  That same picture requires that partons from the $p_z<0$
proton have a broad distribution in $x$.  Forward dijets select the
low-$x$ component of that distribution, so could be of interest to
further probe tri-gluon correlator contributions.

\begin{figure}
   \centering
   \includegraphics[width=0.36\textwidth,clip]{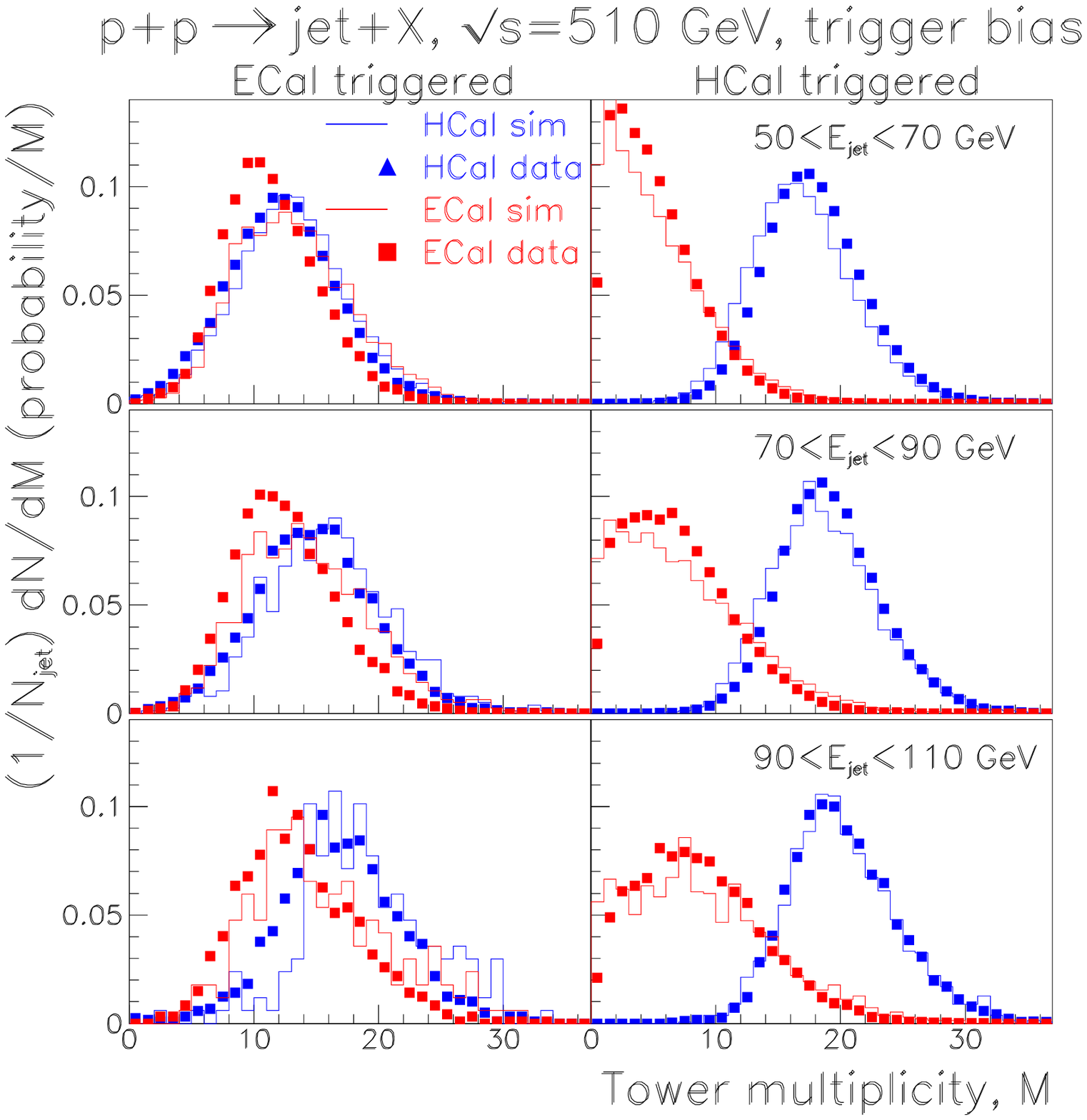}
   \includegraphics[width=0.36\textwidth,clip]{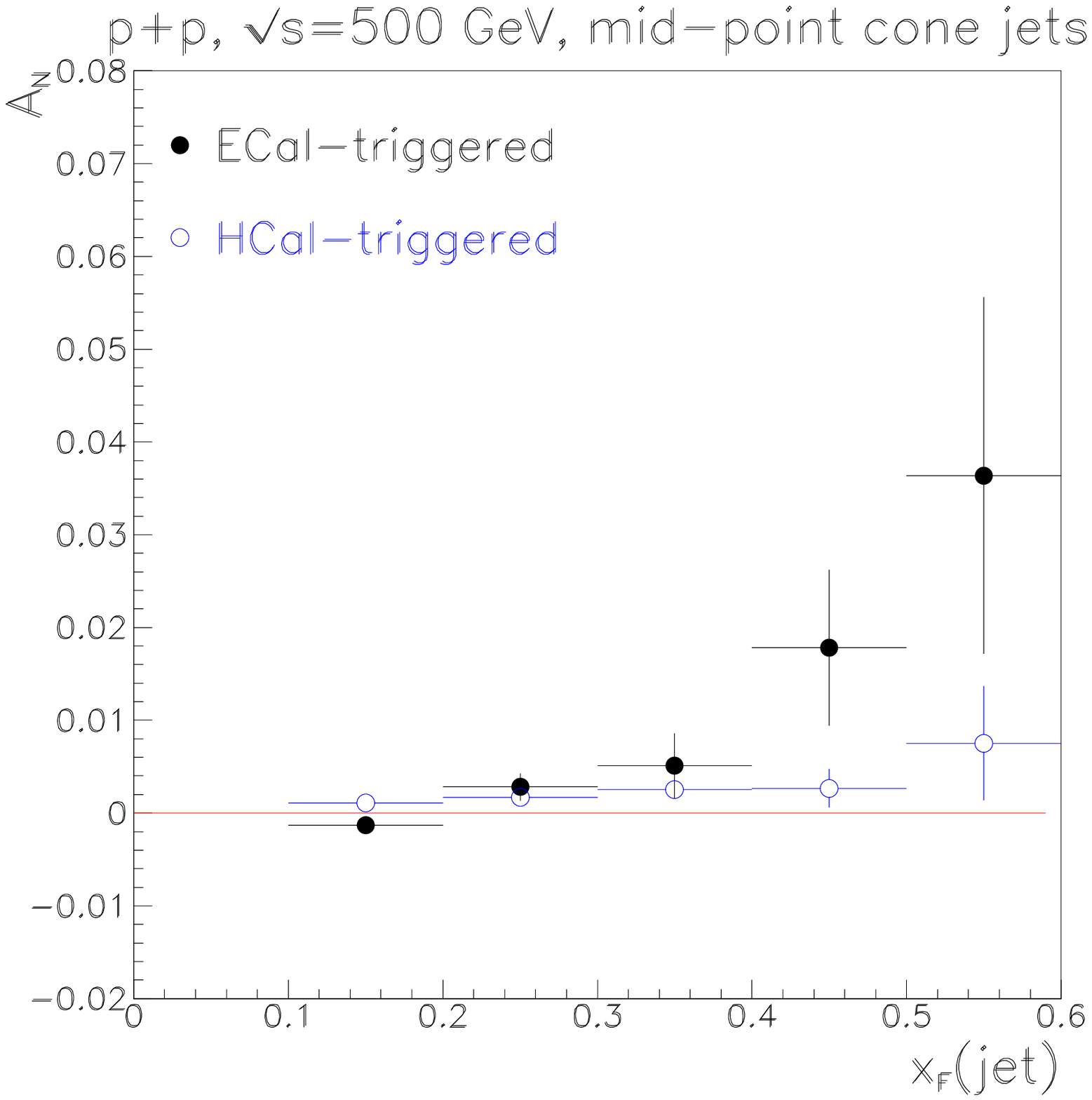}
   \caption{The impact of trigger bias is evident for mid-point cone
   jets reconstructed for events with an ECal trigger, corresponding
   to $\sum E_{ECal}\geq22$ GeV. (Top) tower
   multiplicity distributions show a bias towards jets with larger
   electromagnetic multiplicities for an ECal trigger; and (Bottom) the
   trigger bias makes jets more $\pi^0$ like, thereby impacting their
   spin asymmetry.}
\label{jet_bias}         
\end{figure}

On the topic of jets, there are two notes of caution.  Jet-finding
algorithms can be applied to any detector response in a given
$\eta-\phi$ acceptance.  I think we should be careful with our
language, in that not all clusters of energy deposition are jets.
Jets should be clustered energy depositions that are related back to
the momentum of scattered partons.  Absent that connection, it is difficult to relate
an experimental observable to an object treated by theory.  The second
caution is in regard to trigger bias, and its impact on reconstructed
jets.  An example of this was obtained from the apparatus in Fig.~\ref{jet_figs}.
That apparatus was the first stage of what was to become an experiment
that would measure spin observables for DY production \cite{ANDY_PAC}.  As such, there
were small electromagnetic calorimeter (ECal) modules.  The ECal modules were
used to trigger (via a sum of ADC values from all cells of each ECal
module, corresponding after final calibrations to
$\sum E_{ECal}\geq22$ GeV) readout for a small sample of events obtained in
$p^{\uparrow}p$ collisions at $\sqrt{s}=500$ GeV. Jets from that data
sample were reconstructed and compared to jets reconstructed from
HCal-triggered readout.  Tower-multiplicity distributions are shown in
Fig.~\ref{jet_bias}.  Evident in that figure is that jets triggered by
the ECal modules bias the fragmentation.  The bias extends well beyond
the $\approx 22$ GeV ECal-trigger threshold.  Comparing transverse SSA
in the right panel of the figure shows that the bias impacts the spin
observable, most likely because the jets include high-energy neutral
pions as selected by the ECal-trigger.

\section{Conclusions and Outlook}

My conclusions will be brief and my outlook will be long, because
there remains much to learn from $p^{\uparrow}p$ collisions at RHIC.

In conclusion, RHIC has clearly demonstrated that
$p^{\uparrow}p\rightarrow\pi X$ with large $x_F$ has large transverse
single spin asymmetries at very high collision energies.  Small and
positive $A_N$ is measured for forward jet production, in similar kinematics.  Unpolarized
$\pi$ and jet cross sections are in agreement with NLO pQCD in the same kinematics,
consistent with a partonic scattering origin to the spin effects.
Most aspects of the measurements can be accounted for by theory, and
suggest a role played by the Sivers effect.

The concurrence of the RHIC results with measurements of transverse
SSA in SIDIS has led to a significant change in how we view the
structure of the proton.  The ideas for the importance of spin-orbit
correlations that were introduced to explain large transverse SSA in
$p^{\uparrow}p\rightarrow\pi X$ at lower $\sqrt{s}$ have been fully
developed.  Phenomenology now talks about orbitting partons as
potentially an important contribution to the spin of the proton,
although much work remains to prove this.

There is a consensus that polarized Drell-Yan production
($p^{\uparrow}p\rightarrow\gamma^* X$ and $\pi p^{\uparrow}\rightarrow\gamma* X$)
is a critical experiment to test a theoretical prediction that the
Sivers function changes sign for polarized DY relative to SIDIS.
The COMPASS collaboration will begin a polarized DY experiment later
this year \cite{Chiosso}. There are proposals to pursue polarized DY
production at many laboratories, as also described at this workshop \cite{Peng}.
RHIC remains the only facility with polarized proton beams, and
remains the world's first and only polarized proton collider.  It is
natural to exploit this uniqueness to address the physics question
regarding the sign change of the Sivers function.  The issues to be
aware of include the precision to which we presently know the Sivers
function from SIDIS and whether we have sufficient understanding of
how the Sivers function evolves with resolution scale.  Most polarized
DY measurements will require $M_{\gamma^*}>4$ GeV/c$^2$ (as set by
background considerations), corresponding to a resolution scale of 16
GeV$^2$, whereas the SIDIS measurements have $<Q^2>\approx2.4$ GeV$^2$
\cite{HERMES-2} and 3.8 GeV$^2$ \cite{COMPASS-1}. 

To meet the requirements for a robust test of the theoretical
prediction at RHIC, forward detection of dileptons from polarized DY
production is essential, so as to match the kinematics of SIDIS as
closely as possible.  The forward produced virtual photon should have
$0.02\leq x_{F,\gamma^*} \leq 0.3$, since in the forward region
$x_{F,\gamma^*}$ is to a very good approximation the Bjorken $x$ of
the quark from the polarized proton.  The $\overline{q}$ from the
polarized proton has Bjorken $x_2\approx
M_{\gamma^*}^2/(x_{F,\gamma^*}s)$, to a very good approximation.  The
$\sqrt{s}=500$ GeV collision energy means $x_2 \approx 2\times
10^{-4}$ for $M_{\gamma^*}$=4 GeV and $x_{F,\gamma^*}$=0.3.  The high
energy of the collider results in large partonic luminosity, to partly
overcome the nucleon-nucleon luminosity advantage of fixed-target experiments.
Estimates of backgrounds were made for a forward calorimeter system
with tracking detectors that would observe $e^+ e^-$ dileptons from
the virtual photon, with the conclusion that backgrounds can be
reduced to $<10$\% of the virtual photon signal.  The measurement
consists of $e/\gamma/$hadron discrimination by differences of their
interactions in matter, as they shower in a calorimeter system.  A
preshower detector before an electromagnetic calorimeter (ECal) and a hadron
calorimeter after the ECal are the primary tools to suppress
backgrounds.  The proposal to make this a specific experiment at IP2
at RHIC was not implemented, so that the interaction region could be
used for a coherent electron cooling experiment.

There is a proposal to implement this concept at STAR, as described at
this workshop \cite{Vossen}.  That proposal includes design and
construction of new forward calorimetry, so likely would not be available
for a polarized DY experiment prior to 2020.

\begin{figure}
   \centering
   \includegraphics[width=0.42\textwidth,clip]{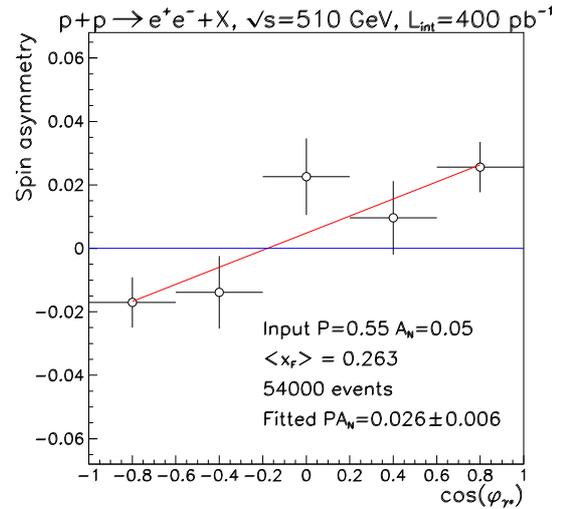}
   \caption{Projected sensitivity to $A_N$ for DY production for a
   forward detection system proposed for installation at STAR in
   2016.  $M_{\gamma^*}>4$ GeV/c$^2$ is imposed, but otherwise
   DY kinematics match those from SIDIS \cite{HERMES-1,COMPASS-2}.}
\label{ANDY_IP6}         
\end{figure}

An implementation of the concept developed for the dedicated
experiment can be made at STAR for 2016, using an existing calorimeter
that could be modified to provide an ECal as the primary tool to detect
di-electrons and an HCal behind it to reject backgrounds.  As had been
proposed, this calorimeter system would include a preshower detector
(whose construction is underway) and tracking detectors.  A test of
this calorimeter was done in the 2014 RHIC run, which included
$^3$He+Au collisions at $\sqrt{s_{NN}}=200$ GeV.  The calorimeter proved
to be robust against challenging beam conditions, as is a requirement
for $p^{\uparrow}p\rightarrow\gamma^* X$.  Lead glass operated during
the earlier $W$ physics program was badly discolored by radiation
damage, so does not appear suitable for a forward DY experiment.  The
bottom line is that a path exists for a polarized DY experiment to
begin at STAR in 2016.  Many steps remain before this path is approved
and a forward DY experiment at STAR is completed.  Projected
statistical undcertainty for measuring $A_N$ for forward DY production
is shown in Fig.~\ref{ANDY_IP6}.  The kinematics is chosen to match
those of SIDIS, except that $M_{\gamma^*}>4$ GeV/c$^2$.

Since forward DY may be pursued with a calorimetric apparatus, there
are other tantalizing prospects for transverse spin physics on the
horizon.  Most notably, is jet physics, where $\pi^0$ within the jet
can be accessed.  A robust measurement can help to establish the
fragmentation contribution to $p^{\uparrow}p\rightarrow\pi X$.  In
addition, the calorimeteric system for polarized DY in the forward
direction looks promising for reconstruction of $\Lambda$ \cite{Bl13},
although discrimination of $\overline{\Lambda}$ is difficult.  This
opens the prospects for a measurement of induced polarization at
large $x_F$ and for a measurement of polarization transfer ($D_{NN}$)
for $p^{\uparrow}p\rightarrow\Lambda^{\uparrow}X$ at $\sqrt{s}=500$ GeV.

A bright future for continued polarized proton operations at RHIC is
on the horizon.  Realization of that future is the goal.

\section*{Acknowledgements}
Many people have been involved in the RHIC-spin program.  Special
thanks to the Collider-Accelerator department, the STAR collaboration
and the A$_N$DY collaboration.  I am extremely grateful for my close collaboration
with H.J.~Crawford, J.~Engelage, A.~Ogawa, and L.~Nogach.

\end{document}